\documentclass[prl,twocolumn,superscriptaddress,floatfix,showpacs]{revtex4}
\usepackage{graphicx,amsfonts,amssymb,amsmath, hyperref}

\newif\ifhyper
\hypertrue
\ifhyper
\hypersetup{
   citecolor = {green},
   colorlinks = {true}, 
   urlcolor = {blue} 
}
\fi

\newcommand{\beq}{\begin{equation}}
\newcommand{\eeq}{\end{equation}}
\newcommand{\beqa}{\begin{eqnarray}}
\newcommand{\eeqa}{\end{eqnarray}}

\def\Longarrow{\protect\@lra}
\def\@lra{\relbar\joinrel\relbar\joinrel\relbar\joinrel%
          \relbar\joinrel\rightarrow}

\begin{document}

\title{Spin-$S$ Kagome quantum antiferromagnets in a field with tensor networks}

\author{Thibaut Picot}
\affiliation{Laboratoire de Physique Th\'eorique, IRSAMC, CNRS and Universit\'e de Toulouse, UPS, F-31062 Toulouse, France}

\author{Marc Ziegler}
\affiliation{Institute of Physics, Johannes Gutenberg University, 55099 Mainz, Germany}

\author{Rom\'an Or\'us}
\affiliation{Institute of Physics, Johannes Gutenberg University, 55099 Mainz, Germany}

\author{Didier Poilblanc} 
\affiliation{Laboratoire de Physique Th\'eorique, IRSAMC, CNRS and Universit\'e de Toulouse, UPS, F-31062 Toulouse, France}

\begin{abstract}

Spin-$S$ Heisenberg quantum antiferromagnets on the Kagome lattice offer, when placed
in a magnetic field, a fantastic playground to observe exotic phases of matter with
(magnetic analogs of) superfluid, charge, bond or nematic orders, or a coexistence of several of the latter.  
In this context, we have obtained the (zero temperature) phase diagrams up to $S=2$ directly in the thermodynamic limit thanks to infinite Projected Entangled Pair States (iPEPS), 
a tensor network numerical tool. 
We find incompressible phases characterized by a magnetization plateau vs field
and stabilized by spontaneous breaking of point group or lattice translation symmetry(ies). The nature of such phases may be semi-classical, as the plateaus at  $\frac{1}{3}$th, $(1-\frac{2}{9S})$th and $(1-\frac{1}{9S})$th of the saturated magnetization (the latter followed by a macroscopic magnetization jump), or fully quantum as
the spin-$\frac{1}{2}$ $\frac{1}{9}$-plateau exhibiting  coexistence of charge and bond orders. Upon restoration of the spin rotation $U(1)$ symmetry a finite compressibility appears, although lattice symmetry breaking persists. For integer spin values we also identify spin gapped phases at low enough
field, such as the $S=2$ (topologically trivial) spin liquid with no symmetry breaking, neither spin nor lattice. 
\end{abstract}

\maketitle

\emph{Introduction.-}
The antiferromagnetic quantum Heisenberg model on the Kagome lattice (KHAF) is one of the most intriguing strongly-correlated systems. Because of the geometric frustration of the lattice, the spin-$\frac{1}{2}$ case ($S=\frac{1}{2}$) has proven exceptionally hard to understand: after more than twenty years of study, there is still no clear consensus about the nature of its ground state \cite{s12_1}, though recent Density Matrix Renormalisation Group~\cite{s12sl} and Variational Monte Carlo~\cite{s12csl} computations suggest the emergence of a topological or critical quantum spin liquid (SL), respectively.
Tensor network numerical tools have come up with
good variational energies for the KHAF~\cite{xie}. In the presence of a magnetic field the system exhibits several incompressible phases detectable as plateaus in the longitudinal magnetization~\cite{plat12,nishimoto_spin12}.  An external magnetic field may also give rise to additional plateaus which behave like chiral spin liquids in the XY regime~\cite{frad}. The model has also been studied for larger values of the spin, for which the strong quantum fluctuations present in the $S=\frac{1}{2}$ case are weaker. For instance, for spin $S=1$ the ground state (GS) is believed to be a simplex solid \cite{spin1,spin1pd} (yet other alternatives have also been put forward \cite{go, spin1other,hss}), and in the presence of a field a rich phase diagram was predicted including nematic and supernematic phases \cite{spin1pd}. It is also suspected that incompressible phases for non-zero field persist for larger spin values \cite{spinlarge}. The KHAF is also of experimental relevance, e.g., minerals such as herbersmithite ZnCu$_3$(OH)$_6$Cl$_2$, volborthite Cu$_3$V$_2$O$_7$(OH)$_2\cdot 2$H$_2$O and vesignieite BaCu$_3$(OH)$_6$Cl$_2$ can be described by the $S=\frac{1}{2}$ KHAF \cite{her}. Some nickelate or vanadate compounds such as KV$_3$Ge$_2$O$_9$ and BaNi$_3$(OH)$_2$(VO$_4$)$_2$ consist also of weakly coupled $S=1$ KHAF layers \cite{van}, and  minerals such as chromium-jarosite KCr$_3$(OH)$_6$(SO$_4$)$_2$ have a KHAF structure with spin $S=\frac{3}{2}$ \cite{spinlarge}. 

In fact, the KHAF is a fantastic toy model to study the crossover between quantum and classical mechanics. This is so  because for the smallest-possible quantum spin value, i.e.,  $S=\frac{1}{2}$, quantum fluctuations due to geometric frustration are so strong that the GS at zero field stabilizes, to our best evidence so far, in a quantum spin liquid, which is a inherently quantum-mechanical phase. However, as the spin gets larger, quantum fluctuations get weaker as the classical limit $S \rightarrow \infty$ is approached.
Understanding this quantum-to-classical cross-over would throw some more light on how quantum effects can be enhanced or suppressed in quantum lattice systems. In turn, this may also provide guidance in the search of new exotic phases of quantum matter, and novel quantum materials.  

Motivated by this, here we analyze the zero-temperature phase diagram of the KHAF up to spin $S=2$ with a magnetic field. For the numerical simulations we use tensor network methods \cite{tn} based on infinite Projected Entangled Pair States (iPEPS) \cite{iPEPS,jiang}, which work directly in the thermodynamic limit. In the presence of a field we find a host of quantum phases of matter at zero temperature for different spin values. These phases may correspond to a magnetization plateau (incompressible) or not (compressible), and also break spontaneously or not $U(1)$ (spin rotation), translation, $C_6$ and/or reflexion symmetries, see Table \ref{tab:symmetries}. 
The associated long range orders are the magnetic analogs of the electronic
charge density (here named ``Solid"), bond density (here named ``Valence Bond Crystal'' or VBC), nematic and superfluid orders. 
Incompressible phases, corresponding to plateaus, shrink quickly -- or disappear -- when the value $S$ of the spin increases. Moreover, we also find macroscopic magnetization jumps at large fields, in accordance with a description of independent magnons.\cite{magnons} 

\emph{Model and methods.-} 
Here we consider the antiferromagnetic quantum Heisenberg model in the presence of an external magnetic field $h$ along the $z$ direction on the Kagome lattice (KHAF),
\beq
H = \sum_{\langle i j \rangle} \mathbf{S}_i \cdot \mathbf{S}_j - h \sum_i S_i^z. 
\eeq
We will use values of the spin $S$ up to $S = 2$. Note that spin-rotation around $z$ ($U(1)$ symmetry) is preserved. As discussed above, this model exhibits a rich variety of exotic quantum behaviours for different spin values. 

\begin{figure}
\includegraphics[trim=0mm 60mm 0mm 60mm, clip, width=\linewidth]{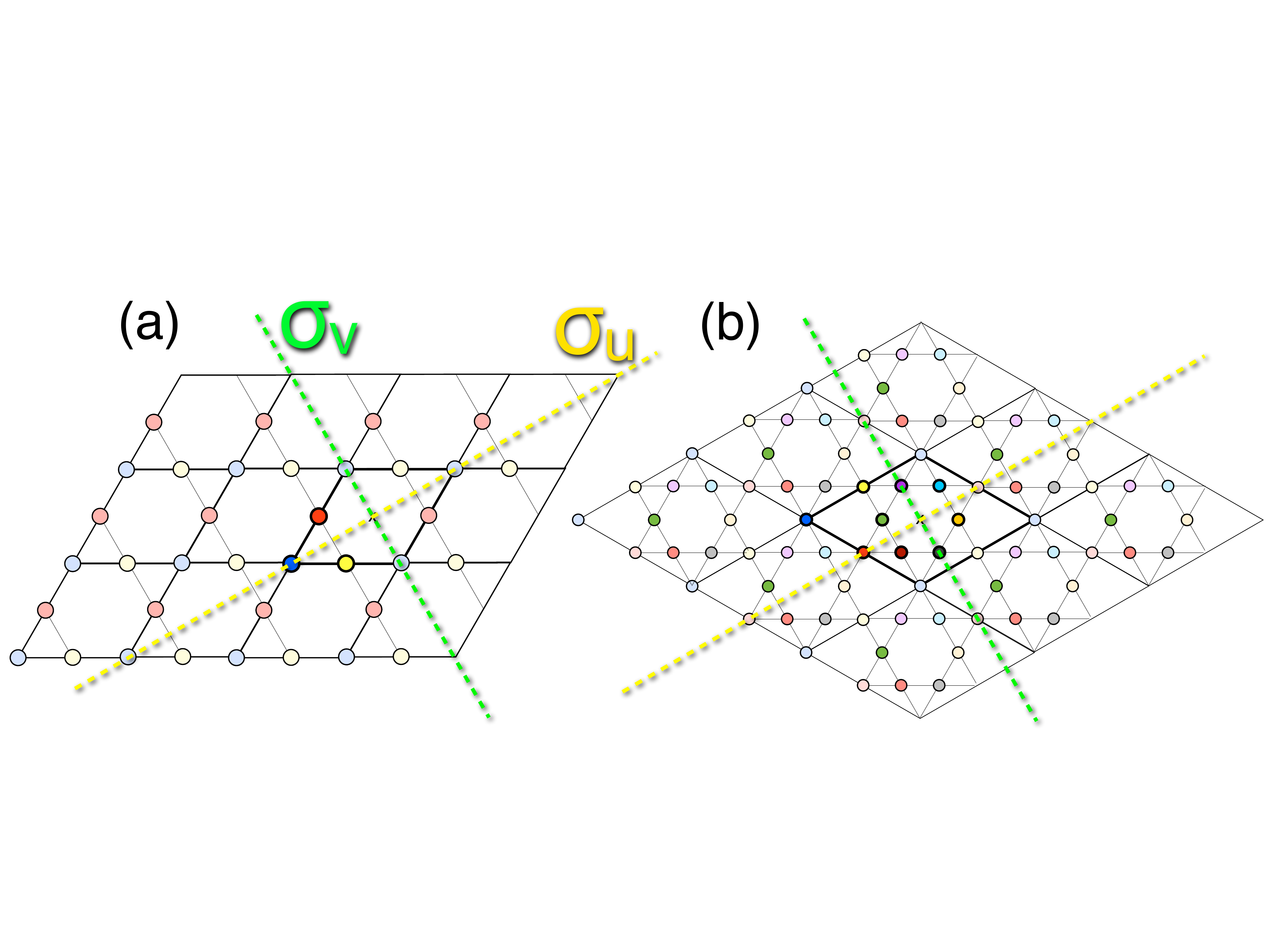}
\includegraphics[trim=0mm 60mm 0mm 80mm, clip, width=\linewidth]{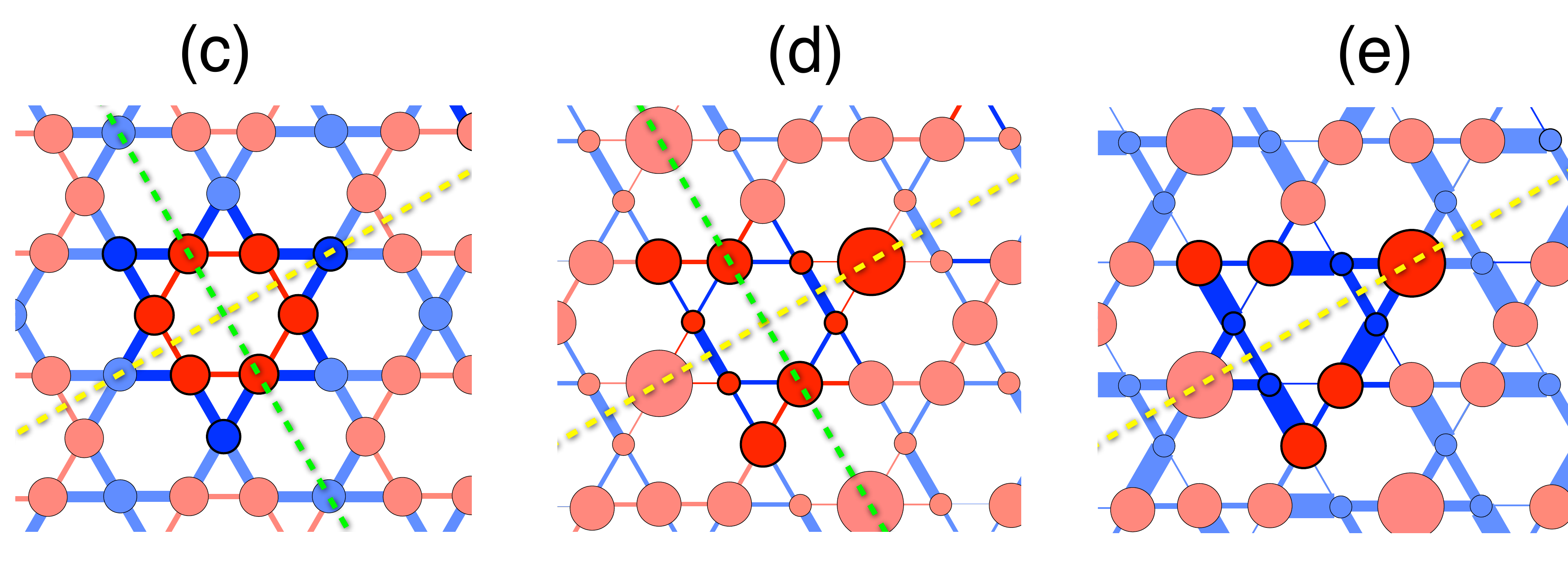}
\caption{[Color online] Two patterns from the 9 sites unit cell calculations which (a) preserves (the 
original 3 sites unit cell is recovered) or (b) breaks translation symmetry ($\sqrt{3}\times\sqrt{3}$ superstructure). In the figure, the $\sigma_u$ and  $\sigma_v$ reflexion symmetry lines are represented by yellow and green dashed lines, respectively.  $\frac{2\pi}{6}$-rotations around the crossing point of the two dashed lines generate a $C_6$ symmetry. We also show the phases (c) Solid 1, (d) Solid 2 and (e) VBC-Solid from Table~\ref{tab:symmetries}. The circles represent the longitudinal magnetization $\langle S^z \rangle$ at every site, and the lines represent the local energy terms $\langle \mathbf{S}_i\cdot\mathbf{S}_{j}\rangle$. In all cases, red (blue) means positive (negative), with the size/thickness proportional to the absolute value. These three plots correspond, in particular, to the three plateaus in the $S=\frac{1}{2}$ phase diagram  at magnetization $\frac{1}{3}$ (c), $\frac{5}{9}$ (d) and $\frac{1}{9}$ (e) in Fig.~\ref{fig:m_z}(a). Notice that the phase (e) is, conceptually, a VBC with coexisting solid order, rather than a pure solid phase. If the valence bonds strongly resonate, the  VBC order ``melts" and one recovers the Solid 2 phase in (d).}
\label{fig:unit_cells}
\end{figure}

\begin{table}
\vspace{1em}
\begin{tabular}{|c||c|c|c|c|c|c|}
  \hline
    & $U(1)$ & $\mathcal{G}_T$ & $C_6$ & $\sigma_u$ & $\sigma_v$ & $d$\\
  \hline
  \hline
 Spin Liquid (SL) & \checkmark & \checkmark & \checkmark & \checkmark & \checkmark & 1\\
  \hline
 Simplex Solid (SiSo)  & \checkmark & \checkmark & $C_3$ & \checkmark & $\times$ & 2\\
  \hline
  Superfluid (SF)  & $\times$ & \checkmark & \checkmark & \checkmark & \checkmark& 1\\
  \hline
  Solid 1 (S1)  & \checkmark & $\sqrt{3}\times\sqrt{3}$ & \checkmark & \checkmark & \checkmark & 3\\
 \hline
  Solid 2 (S2)  & \checkmark & $\sqrt{3}\times\sqrt{3}$ & $C_2$ & \checkmark & \checkmark & 9\\
  \hline
  VBC-Solid (VBCS)  & \checkmark & $\sqrt{3}\times\sqrt{3}$ & $\times$ & \checkmark & $\times$ & 18\\
  \hline
  Nematic (N)  & \checkmark & \checkmark & $C_2$ & \checkmark & \checkmark &3\\
  \hline
  Supersolid 1 (SS1) & $\times$ & $\sqrt{3}\times\sqrt{3}$ & \checkmark & \checkmark & \checkmark & 3\\
  \hline
 Supersolid 2 (SS2) & $\times$ & $\sqrt{3}\times\sqrt{3}$ & $C_2$ & \checkmark & \checkmark & 9\\
  \hline 
Super VBCS (SVBCS) & $\times$ & $\sqrt{3}\times\sqrt{3}$ & $\times$ & \checkmark & $\times$ & 18\\
  \hline
  Supernematic (SN) & $\times$ & \checkmark & $C_2$ & \checkmark & \checkmark & 3\\
  \hline
\end{tabular}
\caption{Comparison between different field-induced phases as characterized by their preserved (\checkmark) or broken ($\times$) 
point group $C_6$, translation group $\mathcal{G}_T$, spin-$U(1)$ and reflexion $\sigma_u$ and $\sigma_v$ symmetries. For the $\mathcal{G}_T$-symmetry breaking, the phase is depicted by 9 sites per unit cell, the so-called $\sqrt{3}\times\sqrt{3}$ phase. The rotation symmetry $C_6$ can be broken either into a lower rotation symmetry group $C_3$ or $C_2$, or completely ($\times$). Parameter $d$ is the GS degeneracy, in accordance with the remaining discrete symmetries.
The spin liquid listed here is topologically trivial (since $d=1$).
When the continuous $U(1)$ symmetry is broken, a zero-energy (Goldstone) mode is expected. 
}
\label{tab:symmetries}
\end{table}

Our numerical calculations are done using tensor network methods based on iPEPS \cite{iPEPS}. More specifically, we run algorithms for imaginary-time evolution in order to obtain approximations of the GS of the system in the thermodynamic limit. As explained in the supplementary material, we use two different algorithms. Algorithm 1 is based on mapping the kagome lattice to the square lattice while keeping the locality of interactions, and applying then the simple update for a square-lattice PEPS. We use such method to estimate the ground-state energy at zero field for the spin one-half case (see also supplementary material). In algorithm 2, we use a simplex representation of the tensor network~\cite{xie} and a simple update scheme as well~\cite{jiang}, and given its efficiency, we use it for the finite-field calculations. The refining parameter of our algorithms is the so-called bond dimension $D$, which controls the amount of entanglement in the tensor network, and which we consider up to $D=15$. In algorithm 2, Our PEPS is built from a unit cell of 9 sites together with 6 simplex tensors. Translation symmetry may
be preserved as in the 3-site pattern of Fig.~\ref{fig:unit_cells}(a) (point group symmetry only is broken), 
or spontaneously broken as in the $\sqrt{3}\times\sqrt{3}$ superstructure, i.e. 9-site pattern of Fig.~\ref{fig:unit_cells}(b). The characterization of the phases is possible by checking the local longitudinal and perpendicular magnetizations as a function of the magnetic field, as well as the local energies $\langle\mathbf{S}_i \cdot \mathbf{S}_j\rangle$ between all nearest sites within the unit-cell. To estimate expectation values of local observables, we consider here an approximate mean-field-like effective environment around the unit cell.

\begin{figure*}
\includegraphics[trim=0 0 0 0,clip ,width=1\textwidth]{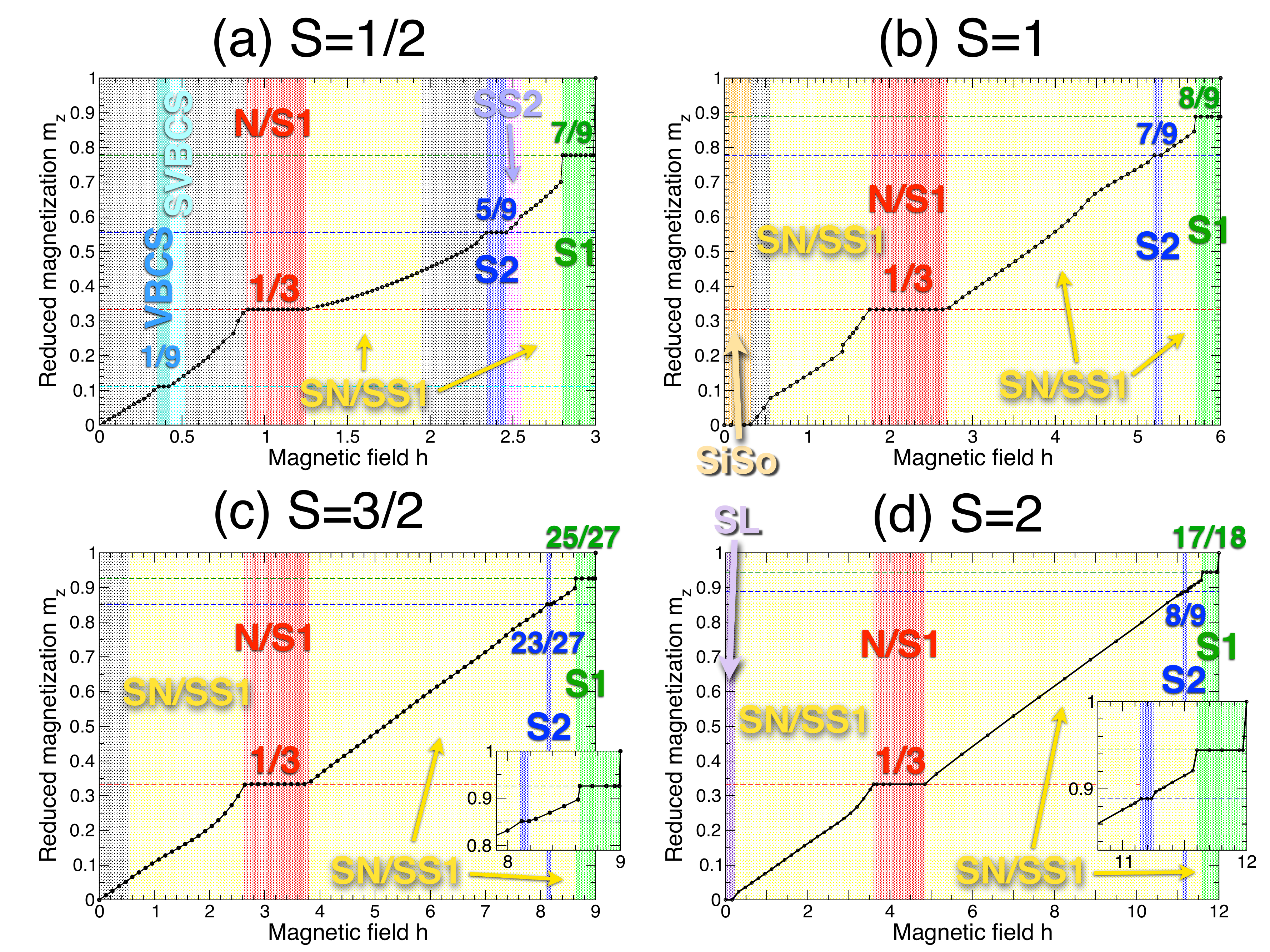}
\caption{
[Color online] Longitudinal magnetization (normalized w.r.t. its value at saturation) versus the external magnetic field for (a) $S=\frac{1}{2}$, (b) $S=1$, (c) $S=\frac{3}{2}$ and (d) $S=2$, with bond dimension $D=10$. The $\frac{1}{3}$-plateau (in red) is present for all spin values, and within our accuracies it is compatible with either a nematic or a solid phase. The dark blue phase, with a commensurable value of the magnetization $(1-\frac{2}{9S})$, corresponds to the 2-magnons phase. The green part corresponds to the 1-magnon phase with magnetization $(1-\frac{1}{9S})$. Both the 1- and 2-magnon states are solid phases. For $S=\frac{1}{2}$, an additional incompressible phase (in light blue) exists at $\frac{1}{9}$ of the total magnetization, which is also a solid phase but with coexisting VBC order (see text). Moreover, for integer spin values a spin gaped phase is identified in brown for low value of the field, with zero magnetization and corresponding to a simplex solid phase for $S=1$ and, to the best of our accuracy, to a topologically trivial spin liquid for $S=2$. Yellow regions are compressible supernematic or supersolid phases. Note that 
a small jump, highlighted by a bond dimension $D=15$, in the $S=1$ magnetization curve (at low field) signals a first order transition between 
two such compressible phases (see Ref.~\onlinecite{spin1pd}). Grey regions could not be fully characterized within our approach. The phases are labelled according to the notations in Table~\ref{tab:symmetries}.}
\label{fig:m_z}
\end{figure*}

\emph{Phase diagrams.-} As we apply an external magnetic field, the system goes through several field-induced phases summarized in Table~\ref{tab:symmetries}. In Figs.~\ref{fig:m_z}(a -- d) we show the computed phase diagrams, presenting the (reduced) longitudinal magnetization as a function of the field for $S=\frac{1}{2}, 1, \frac{3}{2}$ and $2$. In the plots we can see several magnetization plateaus, corresponding to incompressible phases with different symmetry breakings, and where the magnetization attains a commensurate value. Three of these incompressible phases are common to all the local spin values $S$. These are the plateaus at $\frac{1}{3}$th, $(1-\frac{2}{9S})$th and $(1-\frac{1}{9S})$th of the saturated value of the magnetization. In what follows we discuss some of the observed features in these diagrams. 

{(i) $\frac{1}{3}$-plateau:} all the studied values of the local spin show a plateau in the magnetization at one third of the saturation value. 
A full characterization of this phase is subtle. Two competing phases, the Nematic and the Solid 1 phase, are degenerate within the accuracy of our approximations. While both of them are 3-fold degenerate and two sites are always equivalent on every triangle, the Solid 1 phase breaks translation symmetry (see Fig.~\ref{fig:unit_cells}(c)), whereas the Nematic phase breaks the $C_6$ point group symmetry. 

{(ii) 1-magnon $(1-\frac{1}{9S})$-plateau:} this plateau also appears for all values of $S$. It can be understood exactly by a one-magnon picture (see Ref.\cite{magnons}), and has the symmetries of the Solid 1 phase (see Fig.~\ref{fig:unit_cells}(c)). In the $\sqrt{3}\times\sqrt{3}$ superstructure, each {\it independent} magnon (flipped spin in a polarized environment) is localized on an hexagon with $C_6$ symmetry. The one-magnon state is, in fact, also an exact PEPS with bond dimensions $D=6$. The two boundaries of the phase are first order phase transitions, ending in a direct jump to the saturation value.

{(iii) 2-magnon $(1-\frac{2}{9S})$-plateau:} as before, this plateau is also common to all values of the spin $S$. It was first
understood in terms of a $\sqrt{3}\times\sqrt{3}$ superstructure of (quasi-independent) two-magnon states~\cite{magnons}  
(Solid 1). Our numerical optimization, however, shows that the $C_6$ symmetry is lowered down to $C_2$, 
while the two reflexions $\sigma_u$ and $\sigma_v$ are preserved. Such a phase is called Solid 2, according to Table~\ref{tab:symmetries} (see Fig.~\ref{fig:unit_cells}(d)). Although the Solid 1 and Solid 2 phases might be quasi-degenerate, the two-magnon phase with the Solid 1 pattern was not stabilized within our optimization scheme.

{(iv) VBC plateau:} independently of the common incompressible phases of semi-classical origin described above, we also find a specific ``quantum'' plateau for $S = \frac{1}{2}$ at $m_z=\frac{1}{9}$. This phase corresponds to a VBC-Solid (VBCS),  where all symmetries are broken except for the $\sigma_u$-reflexion, thus being $18$-fold degenerate (see Fig.~\ref{fig:unit_cells}(e)). The corresponding pattern can be schematically viewed as polarized spin-$1/2$ forming a $\sqrt{3}\times\sqrt{3}$ superstructure (solid order similar to the S2 phase) superposed with (phase-locked) dimerized 
bow-tie chains running in the $\sigma_u$ direction (VBC order). Note a grand canonical DMRG calculation~\cite{nishimoto_spin12} also revealed this plateau but associated it to a topological $Z_3$ spin liquid with no symmetry breaking, in contrast to our findings.

{(v) Other features:} furthermore, for integer spin values we identify a spin gapped phase at low enough magnetic field. This phase has zero magnetization and corresponds to a simplex solid phase for $S=1$, whereas for $S=2$ it corresponds, to the best of our accuracy,  to a spin liquid with no symmetry breaking. 
We believe the simplest scenario allowed by the Lieb-Schultz-Mattis-Hasting theorem~\cite{lsmh} occurs: the $S=2$ spin liquid is  topologically trivial, its GS being non degenerate. In that case, it belongs to the same class as (and can be adiabatically connected to) the $S=2$ Affleck-Kennedy-Lieb-Tasaki (AKLT) state~\cite{aklt}, a universal resource for measurement-based quantum computation~\cite{qc}, for which we also found a spin gap~\cite{spin1pd}. However, this conclusion disagrees with the coupled cluster calculation~\cite{go} which states that the GS of the spin $S=2$ of the zero-field Heisenberg model has a non-zero magnetization. As shown in Fig.~\ref{fig:m_z}, we also find, for all the spin values, compressible supernematic or supersolid phases which bear the same spontaneous lattice symmetry breaking as their 
nearby solid phases.

\begin{figure}
\includegraphics[trim=0mm 50mm 0mm 50mm, clip, width=1\linewidth]{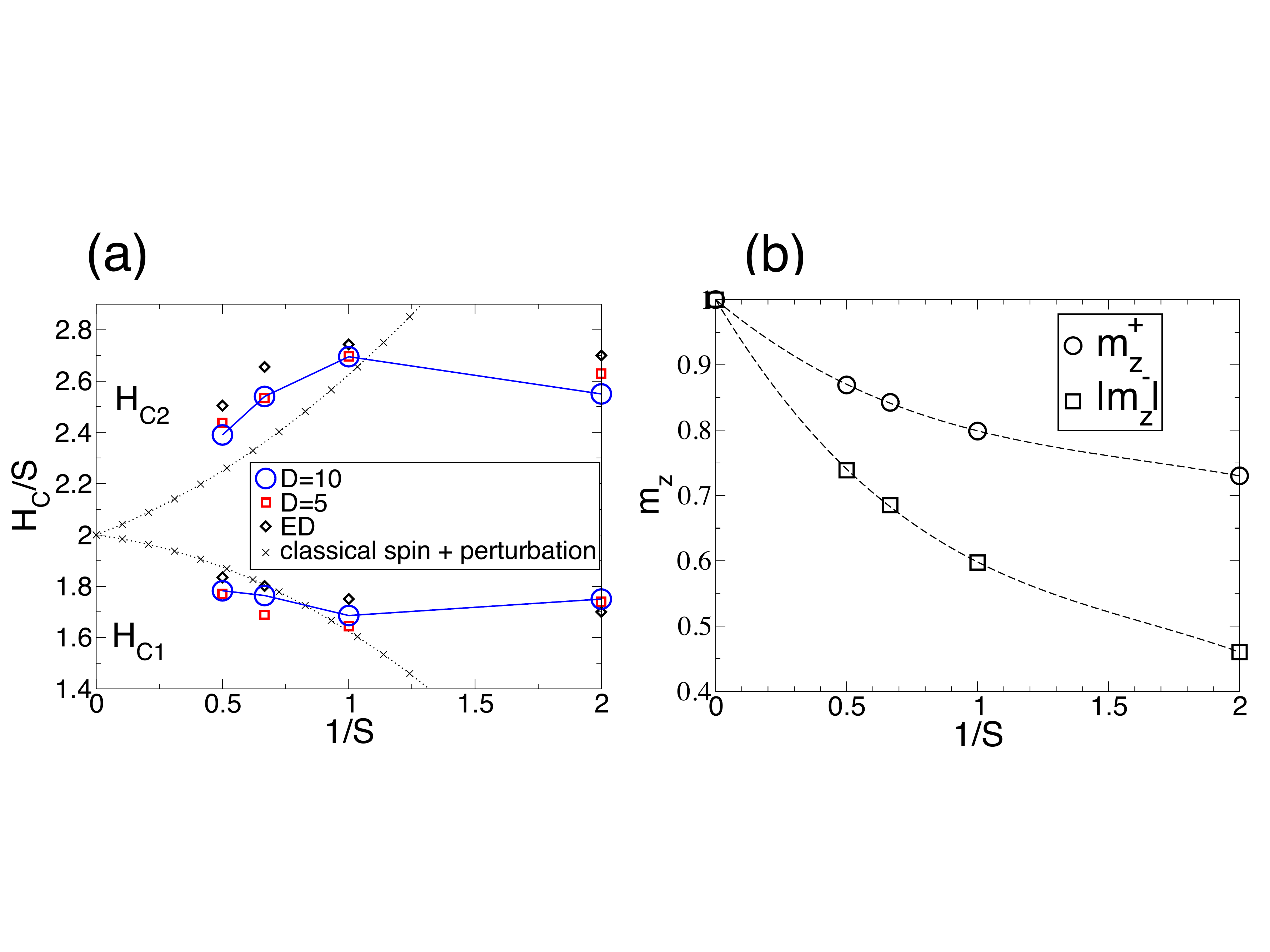}
\caption{[Color online] Properties of the $\frac{1}{3}$-plateau phase. (a) Normalized critical magnetic fields $H_{C1}$ (circles) and $H_{C2}$ (squares) versus the inverse of the spin. Values for bond dimensions $D=10$ (blue) and $D=5$ (red) are shown, as well as results from exact diagonalization and from perturbation theory for classical spins \cite{ed}. (b) Magnitudes of the longitudinal local magnetizations along 
($m_z^+$) and opposite ($m_z^-$) to the field, computed with $D=10$. Dashed lines are polynomial fits in $\frac{1}{S}$.}
\label{fig:plateau_width}
\end{figure}

\emph{Quantum-classical crossover.-} Large values of the spin $S$ tend to quickly suppress quantum fluctuations in the system. In order to study this effect, we have analyzed the rate at which the $\frac{1}{3}$-magnetization plateau disappears as $S$ gets larger. Our results are for bond dimension up to $D=10$ and spin up to $S=2$, and these are presented in Fig.~\ref{fig:plateau_width}. We plot the two critical lines $Hc_1$ and $Hc_2$ in Fig.~\ref{fig:plateau_width}(a). For completeness we also show the results of an exact diagonalization for small clusters from Ref.\cite{ed}. For $S=\infty$, i.e., classical spins, we expect the magnetization curve to become smooth and, therefore, both critical lines end up at the same classical value $Hc_1/S= Hc_2/S= 2$. 
We indeed observe that, moving away from $S=\frac{1}{2}$, the size of the plateau shrinks quickly with increasing $\frac{1}{S}$. 
We also plot the local magnetizations for the $\frac{1}{3}$-plateau versus $\frac{1}{S}$, for $D=10$. As shown in Fig.~\ref{fig:unit_cells}(b), there are two non-equivalent on-site magnetizations, $m_z^+>0$ (in the field direction) and $m_z^-<0$ (opposite to the field direction), such that $\frac{2}{3}m_z^++\frac{1}{3}m_z^-=\frac{1}{3}$. In the classical limit, the quantum fluctuations disappear and $m_z^+=-m_z^-=1$.

\emph{Conclusions and outlook.-} Here we have studied the spin-$S$ KHAF in the presence of a magnetic field up to $S=2$ using tensor network methods based on an iPEPS algorithm and an enlarged supercell. We have characterized a host of (zero temperature) quantum phases, some of semi-classical origin, some purely quantum. Our results confirm and extend in several directions previous results in the literature~:  we show that the well-known $S=1/2$ $m_z=\frac{1}{3}$ and $m_z=1-\frac{1}{9S}$ plateaus extend to larger spin $S$. The spin-$1$ $m_z=0$ simplex solid plateau remains stable within our new enhanced variational space. 
However, we found that the spin-$1/2$ $m_z=1/9$ plateau is stabilized by spontaneous breaking of translation symmetry, in contrast to previous claims.
Also we found a less-symmetric pattern for the $m_z=1-\frac{2}{9S}$ two-magnon plateau than proposed previously. One important finding of the present work is also the first full characterization of all the compressible (i.e. superfluid) phases between the plateaus.

Under completion of this work, we became aware of a related work showing similar findings on the Husimi lattice~\cite{xiang2}. 
 
\acknowledgements
R.O. and M. Z. acknowledge support from the Johannes Gutenberg-Universit\"at, the Deutsche Forschungsgemeinschaft, the MOGON Cluster (Mainz) for CPU time, and discussions with A. Kshetrimayum. T. P. and D. P. acknowledge the NQPTP ANR-0406-01 grant (French Research Council) for support and the CALMIP Hyperion Cluster (Toulouse) for CPU time.
D. P.  also thanks Sylvain Capponi for useful discussions.

\end{document}


\title{Spin-$S$ Kagome quantum antiferromagnets in a field with tensor networks: \textit{supplemental material}}

\author{Thibaut Picot}
\affiliation{Laboratoire de Physique Th\'eorique, IRSAMC, CNRS and Universit\'e de Toulouse, UPS, F-31062 Toulouse, France}

\author{Marc Ziegler}
\affiliation{Institute of Physics, Johannes Gutenberg University, 55099 Mainz, Germany}

\author{Rom\'an Or\'us}
\affiliation{Institute of Physics, Johannes Gutenberg University, 55099 Mainz, Germany}

\author{Didier Poilblanc} 
\affiliation{Laboratoire de Physique Th\'eorique, IRSAMC, CNRS and Universit\'e de Toulouse, UPS, F-31062 Toulouse, France}

\maketitle

\section{Optimization method: Simple Update}
\setcounter{figure}{0}   

\subsection{Description}

Tensor Networks in 2 dimensions in the thermodynamic limit, called \textit{infinite} Projected Entangled Pair State (iPEPS), can be optimized using 
Full Update or Simple Update schemes. The Full Update is more accurate but requires the calculation of the surrounding system of the unit cell -- called environment -- at every step. The calculation of the environment can be performed using a Matrix Product State (MPS)-based approach~\cite{iPEPS}, a Corner Transfer Matrix Renormalization Group (CTMRG)~\cite{iPEPS} or a Coarse Graining Tensor Renormalization Group (CGTRG) method~\cite{jiang} (or any related technic). In contrast, in the Simple Update, the tensor optimization used in this study, the computation of the environment is not done during the optimization stage.
The resulting optimized tensor may have a lower accuracy, but the method allows to go to higher values of the bond dimension D.

\begin{figure}
\includegraphics[trim=0mm 0mm 0mm 0mm, clip, width=0.7\textwidth]{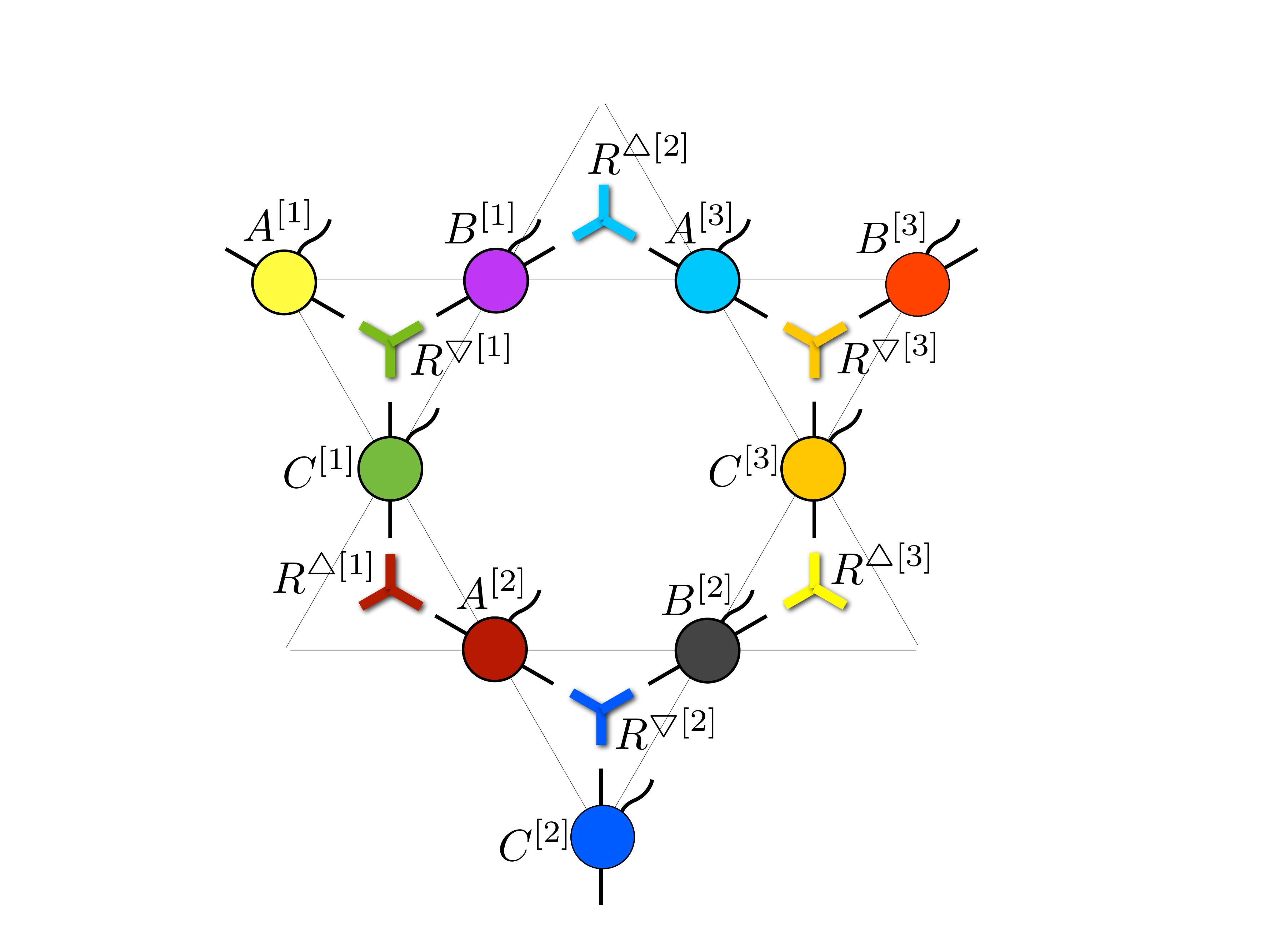}
\caption{Tensor Network of 9 sites per unit cell, composed by 9 site tensors of dimension $dD^2$ and 6 simplex tensors, located on the triangle, of dimension $D^3$.}
\label{fig:9sites_uc}
\end{figure}

\begin{figure}
\includegraphics[trim=47mm 30mm 95mm 40mm, clip, scale=0.5]{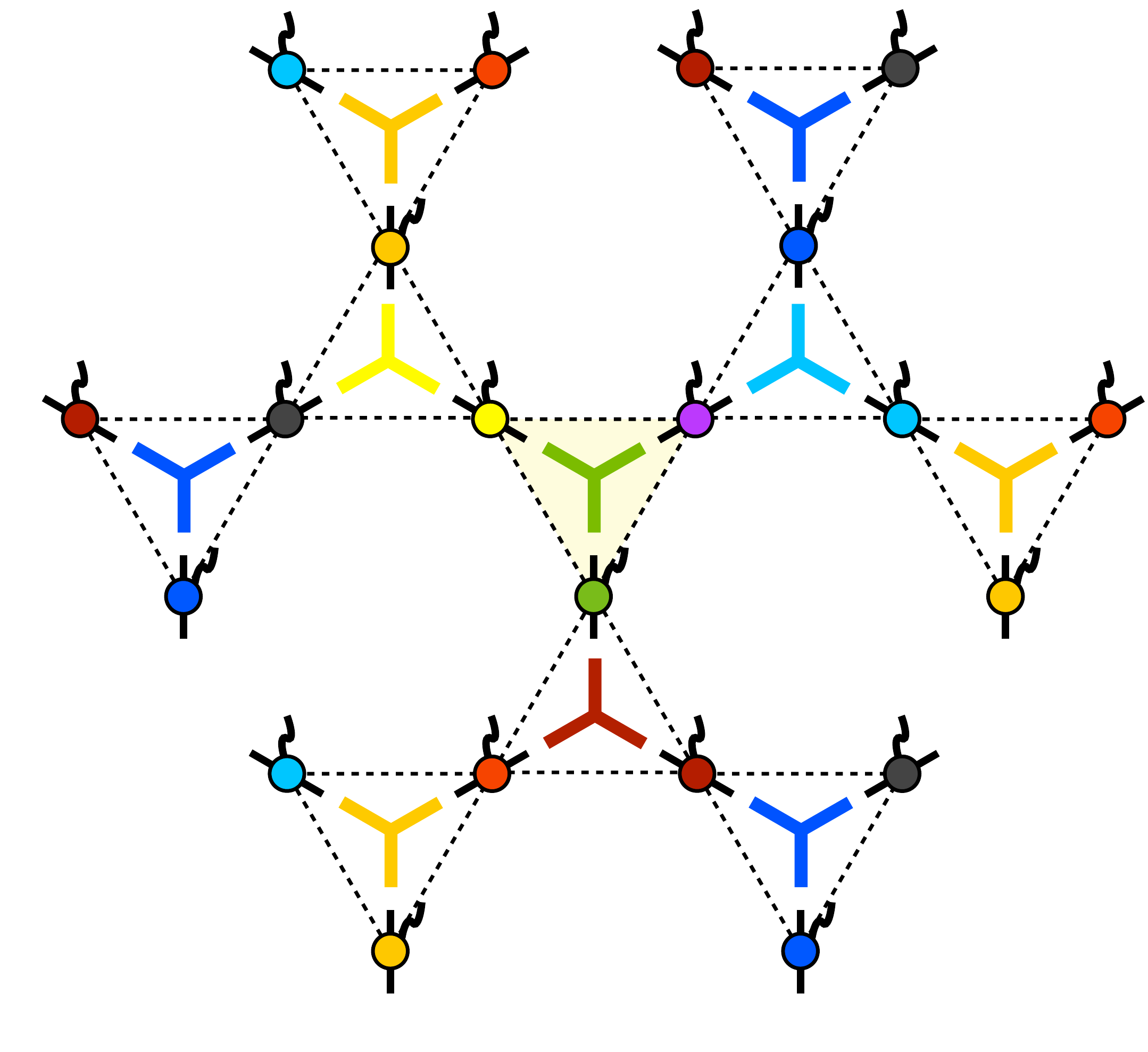}
\caption{Cluster used in the Simple Update approximation. The red, green and blue site tensors picture the A, B and C tensors. The simplex tensors with circle (square) ends depict the down (up) triangle simplex tensors.}
\label{fig:bethe}
\end{figure}

In this paper we use two types of updates. The first one ("algorithm 1") is based on the mapping of the kagome lattice to the square lattice shown in Fig.~\ref{figkagmap}, and is the one used in the calculation of the energy at zero field for the spin-$1/2$ case (see below). In the second one ("algorithm 2"), we use the imbedded cluster of nine sites depicted in Fig.~\ref{fig:9sites_uc}, which is reliable if quantum fluctuations do not extend much beyond the unit cell (it becomes exact if one considers a Bethe lattice). This iPEPS representation considers a large cluster (the focused triangle with its nine neighboring triangles as shown in Fig.~\ref{fig:bethe}), which has proved convenient for the calculations at finite field. Given its efficiency, in this paper we use this type of update to compute the phase diagrams. 

\subsection{Algorithm 1}

\begin{figure}
\includegraphics[trim=0mm 0mm 0mm 0mm, clip, width=0.7\textwidth]{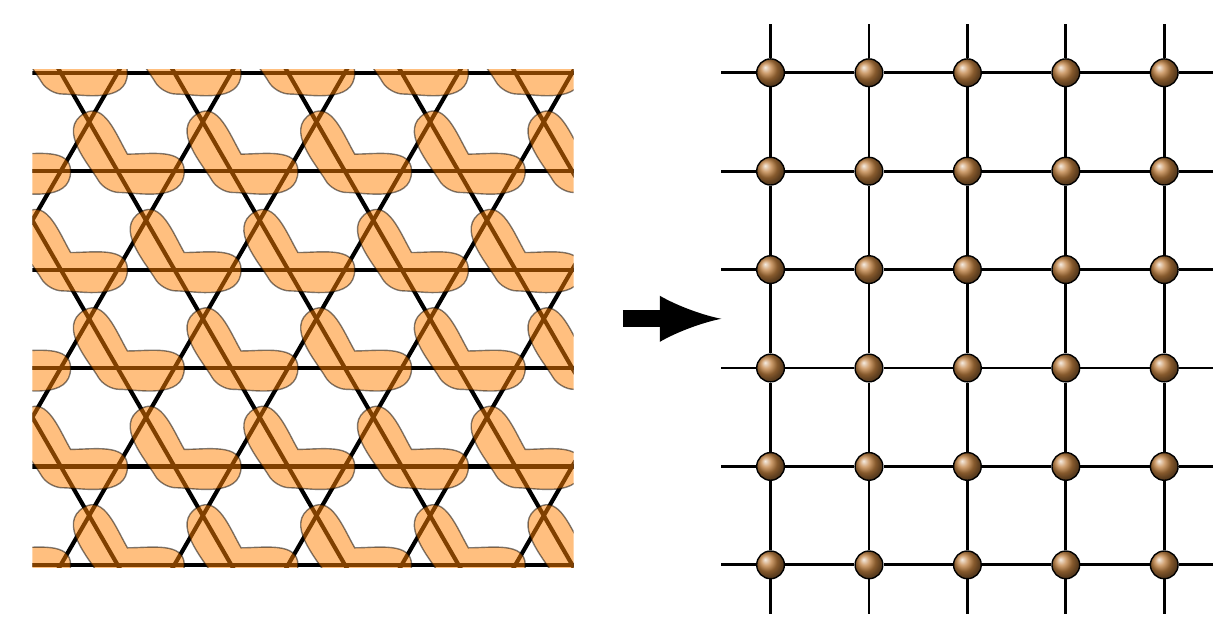}
\caption{Mapping scheme from the kagome to the square lattice. We combine the three orange marked spins together and end up with a square lattice. One single link in the square lattice corresponds to two links in the kagome lattice. Since all nearest neighbours in the kagome lattice are mapped to either one single site or nearest neighbours in the square lattice, the interaction remains local.}
\label{figkagmap}
\end{figure}

The mapping in Fig.~\ref{figkagmap} preserves the locality of interactions in the model (one has, at most, nearest-neighbor interactions also in the square lattice). The system is thus amenable for simulations with a standard iPEPS algorithm on a square lattice, with a squared physical dimension. If a 2-site unit cell is used, then the calculations are, essentially, an improvement upon a 6-site unit cell in the original kagome lattice. The improvement comes from the fact that the three sites that are coarse-grained  are treated by a single tensor instead of three. In our case, here we used also the simple update scheme. 

\subsection{Algorithm 2}

The optimization of the tensors in the Simple Update is based on $\lambda$ matrices which are located on the external bonds of the triangles. In one dimensional system, these $\lambda$ matrices contain the singular values of a decomposition of two neighboring sites, and connect these two sites. In two dimensions, they are an approximation of the outside system, whereas the $R^{\bigtriangledown(\bigtriangleup)}$ simplex tensor connect the three sites inside the focused triangle.

In the following, the algorithm~\cite{xie} for one optimization of the tensors on down triangles around simplex tensor $R^{\bigtriangledown[1]}$ is described (see Fig.~\ref{fig:algo}).
\begin{enumerate}
\item
First, we contract the $T$ tensor on the triangle, as shown in Fig.~\ref{fig:algo}(a),
\[
T^{S_A,S_B,S_C}_{i,j,k}=\sum_{l,m,n}A^{[1]S_A}_{i,l}B^{[1]S_B}_{j,m}C^{[1]S_C}_{k,n}R^{\bigtriangledown [1]}_{l,m,n}
\lambda^{\bigtriangleup A[1]}_i\lambda^{\bigtriangleup B[1]}_j\lambda^{\bigtriangleup C[1]}_k
\]

\begin{figure}
\includegraphics[trim=0mm 90mm 0mm 65mm, clip, width=0.6\textwidth]{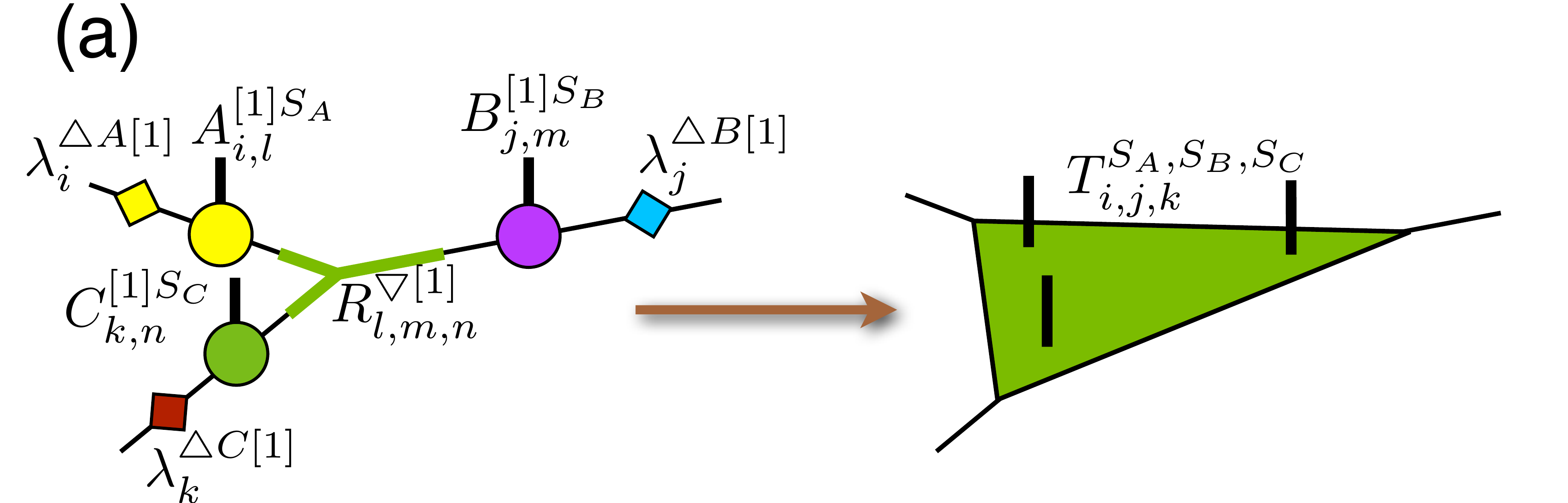}
\includegraphics[trim=0mm 90mm 0mm 65mm, clip, width=0.6\textwidth]{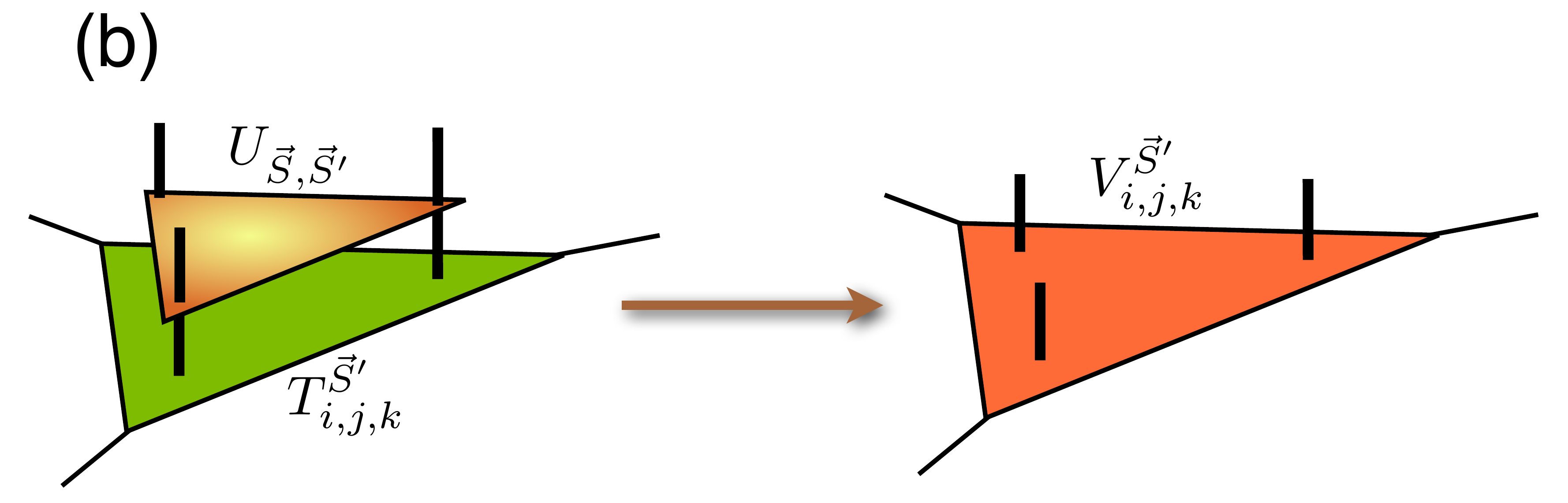}
\includegraphics[trim=0mm 50mm 0mm 10mm, clip, width=0.6\textwidth]{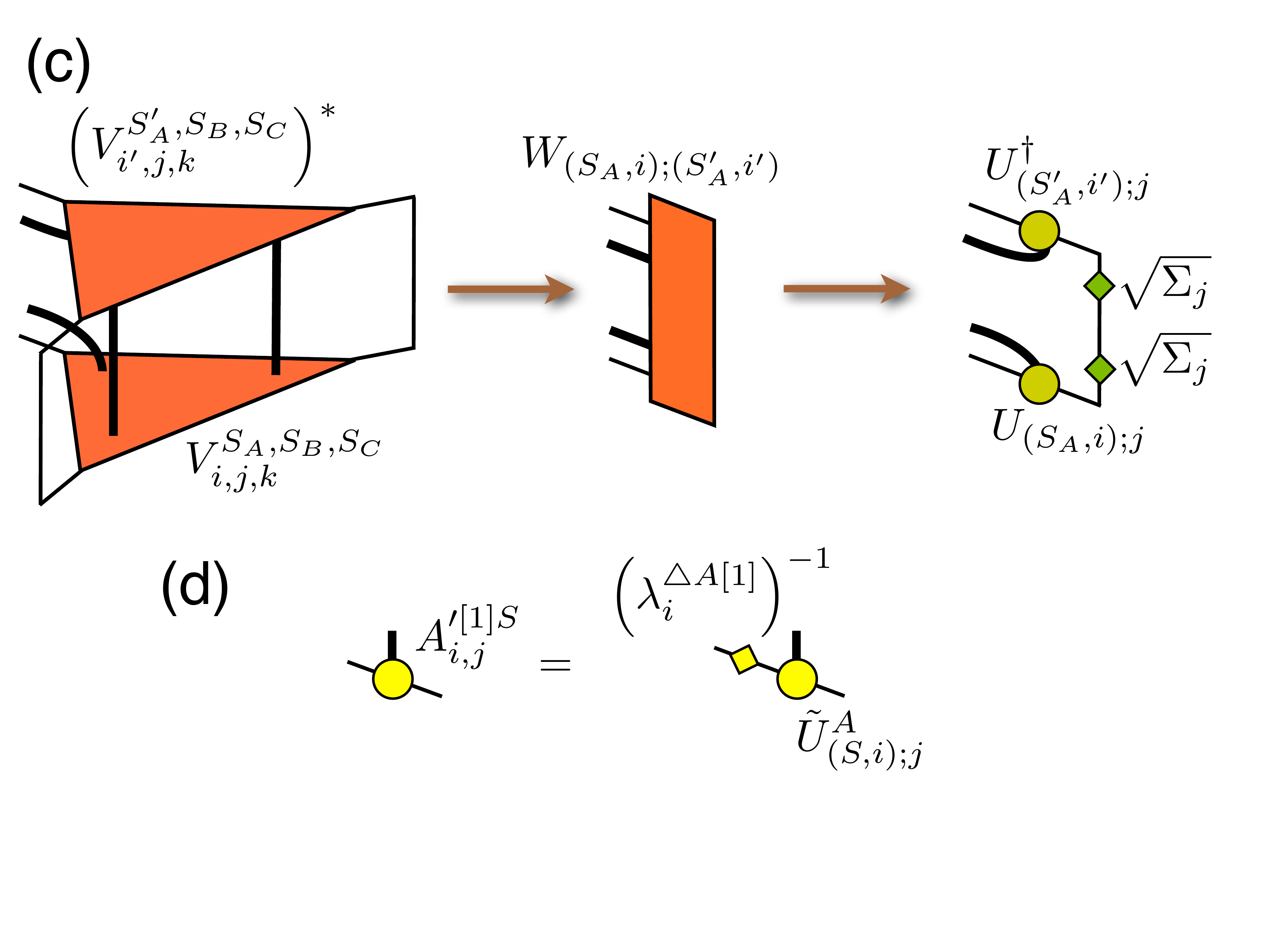}
\caption{$T$ tensor contraction, including $\lambda^{\bigtriangleup}$ diagonal matrices on the down triangle.}
\label{fig:algo}
\end{figure}

\item
Then, we apply the imaginary time evolution operator $U=\exp\left(-\delta \tau H^{\bigtriangledown}\right)$ on $T$: 
\[V^{\vec{S}}_{i,j,k}=\sum_{\vec{S}'}U_{\vec{S},\vec{S}'}T^{\vec{S}'}_{i,j,k}\]
where $\vec{S}$ is a vector, whose components are $S_A$, $S_B$ and $S_C$, and $\delta\tau \ll 1$ (Fig.~\ref{fig:algo}(b)). 

\item
Now, we have to decompose the $V$ tensor into the new tensors $A'^{[1]}$, $B'^{[1]}$, $C'^{[1]}$ and $R'^{\bigtriangledown[1]}$. This step is called a Higher-Order Singular Value Decomposition (HOSVD). We give the example for the $A'^{[1]}$ tensor. We diagonalize the $dD\times dD$ matrix $W=U \Sigma U^\dagger$ as shown in Fig.~\ref{fig:algo}(c).
\[
W_{(S_A,i);(S_A',i')}=\sum_{S_B,S_C,j,k}\left( V^{S_A',S_B,S_C}_{i',j,k}\right)^* \cdot V^{S_A,S_B,S_C}_{i,j,k}
\]

\item
We keep the $D$ largest eigenvalues of the decomposition in $\tilde{\Sigma}^A$, and $\tilde{U}^A$ is the corresponding $dD\times D$ matrix. The new tensors are $A'^{[1]S}_{i,j}=\tilde{U}^A_{(S,i);j}/\lambda^{\bigtriangleup A[1]}_{i}$ (Fig.~\ref{fig:algo}(d)) and $\lambda^{\bigtriangledown A[1]}=\sqrt{\tilde{\Sigma}^A}$ (where $\Sigma ^A \ge 0$ because $W^\dagger=W$). After the updates on the three sites, we can compute the new simplex tensor as \[R'^{\bigtriangledown}_{i,j,k}=\sum_{\{l,m,n\}}\sum_{\{S_A,S_B,S_C\}}T^{S_A,S_B,S_C}_{l,m,n}\left(\tilde{U
}^A_{(S_A,l);i}\right)^*\left(\tilde{U}^B_{(S_B,m);j}\right)^*\left(\tilde{U}^C_{(S_C,n);k}\right)^*\]

\end{enumerate}

The renormalization on up triangles is identical, by doing the change $\bigtriangleup \Longleftrightarrow \bigtriangledown$, and taking $\left(A^{[2]S_A}\right)^T$, $\left(B^{[3]S_B}\right)^T$ and $\left(C^{[1]S_C}\right)^T$. Then, we perform two cyclic permutations on the labels $[1\rightarrow2\rightarrow3]$ in order to optimize all the nine site tensors and the six simplex tensors.
In order to reach the ground state, one can use a succession of several time steps from $\delta\tau = 10^{-1},10^{-2},10^{-3},10^{-4},10^{-5}$. For each time step, one can choose a accuracy criterion for which the evolution is stopped. In addition, one can stop the evolution if the number of time step is too large. Here, we used the normalization of the lambda matrices with a relative accuracy $\Delta=10^{-10}$. When the number of time steps was too large, we stopped the imaginary time evolution at 4000 iterations. In those cases, the order of the accuracy was $10^{-8}$.

\section{Energies}
\setcounter{figure}{0}   

\begin{figure}
\includegraphics[trim=0mm 0mm 0mm 0mm, clip, width=0.7\textwidth]{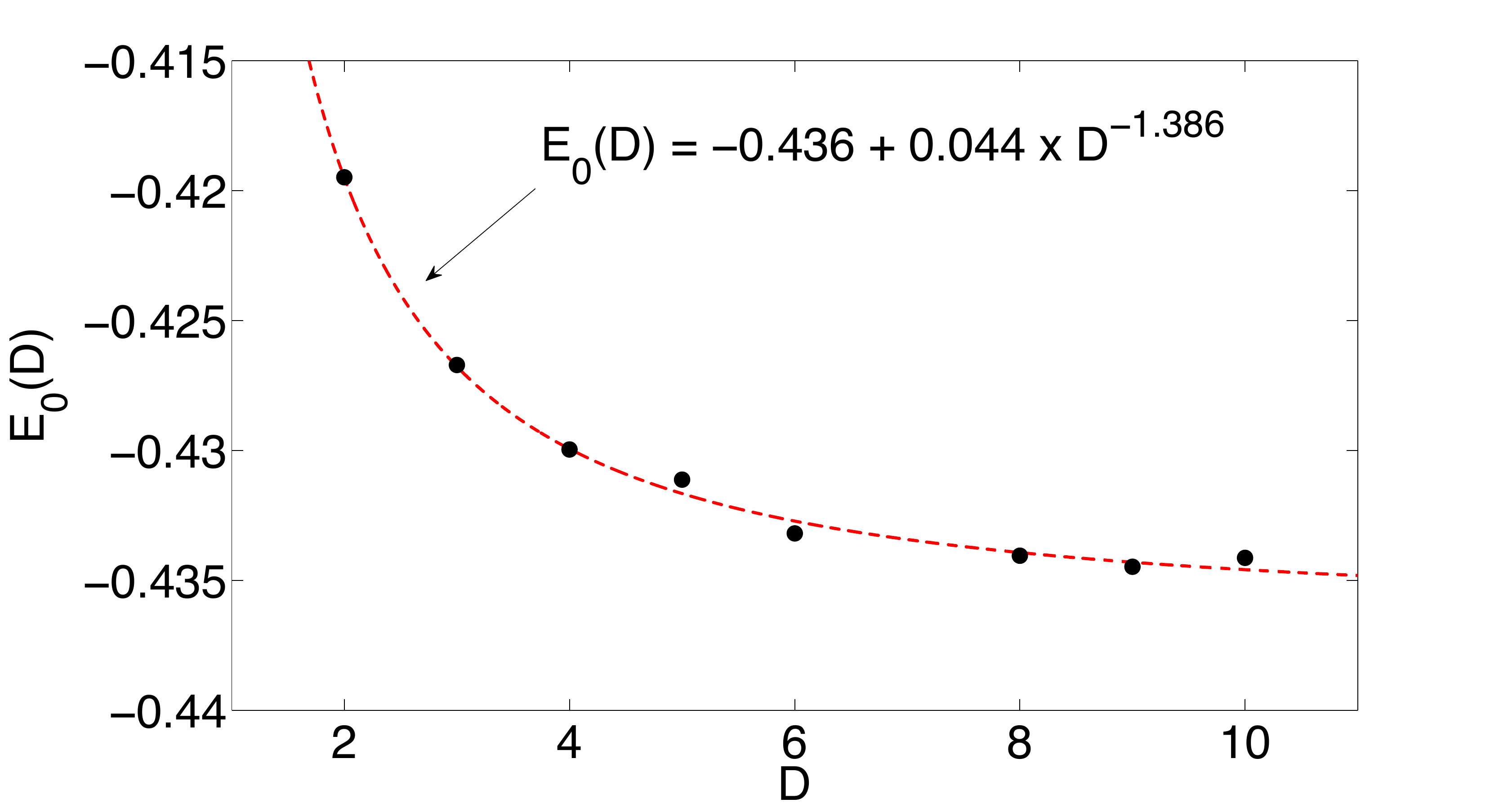}
\caption{Energy of the spin-1/2 case, h=0, for a PEPS with a 6-site unit cell in the kagome lattice, versus the bond dimension $D$.}
\label{fig:energies}
\end{figure}

Here we show some results for the ground state energy for spin $1/2$ and zero field using algorithm 2 above. Our calculations in this case are for a PEPS with a 6-site unit cell in the kagome lattice, where effective environments are computed by Corner Transfer Matrix (CTM) methods, up to PEPS bond dimension $D = 10$. The results are shown in Fig.\ref{fig:energies}, where the points are converged in the environment bond dimension. The dashed line corresponds to an algebraic fit, leading to asymptotic ground state energy $E_0(D\rightarrow \infty) \sim -0.436$. Our numbers obtained using this approach are, in fact, very close to the ones obtained using the PESS approach, which is $E_0^{PESS} = -0.4364$ for bond dimension $D=13$ and a 9-site unit cell~\cite{xie}. 

\begin{figure}
\includegraphics[trim=0mm 40mm 0mm 40mm, clip, width=\textwidth]{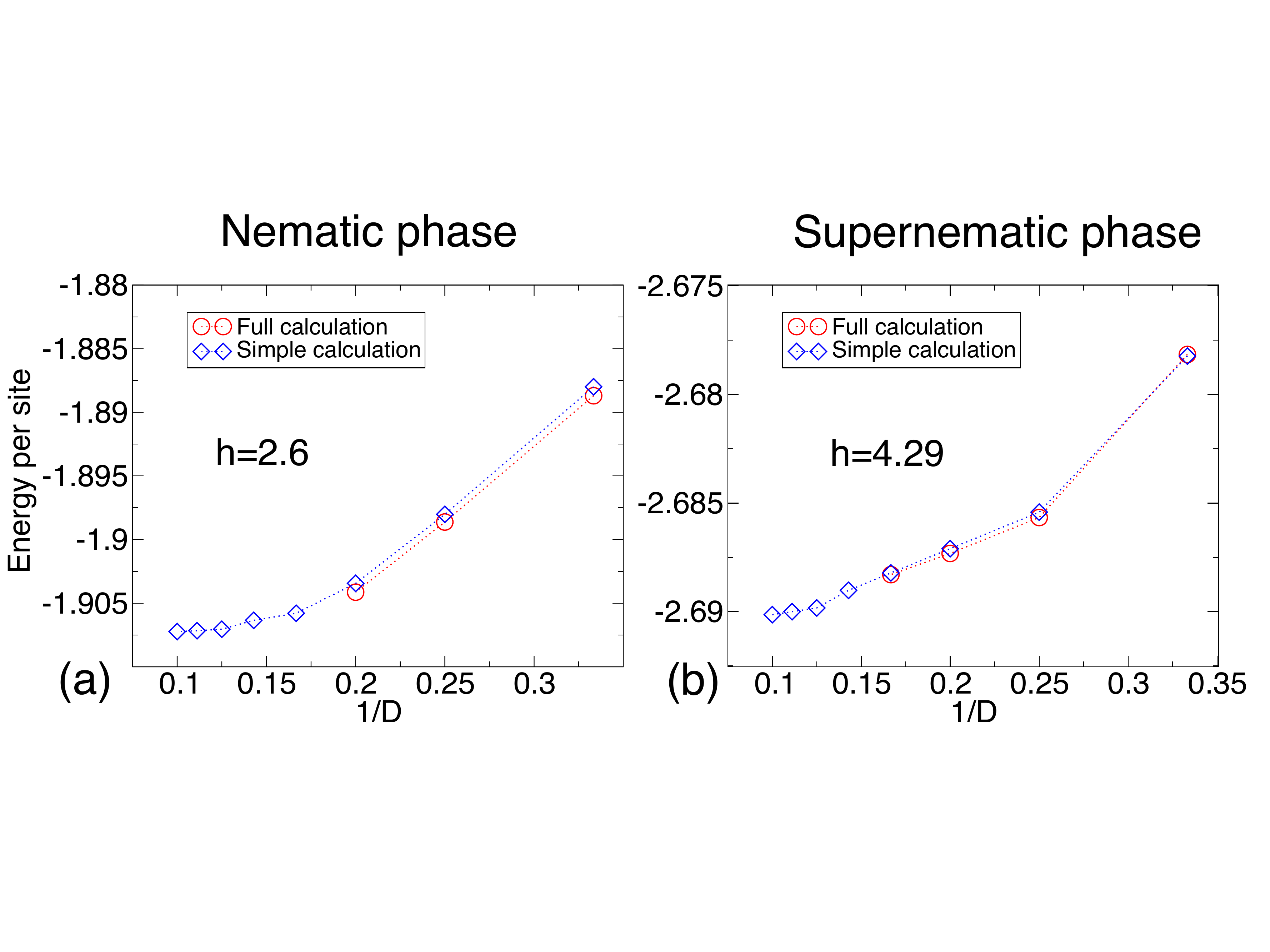}
\caption{Comparison between the simple and the full calculation of the spin-1 case for the nematic (a) and the supernematic phase (b).}
\label{fig:energy_comparison}
\end{figure}

At finite field, we can show that, using algorithm 1 above, the expectation value calculation using only the lambda matrices (called \textit{simple calculation}) is a good approximation compared to the \textit{full calculation}, which corresponds to the calculation of the full surrounding environment. We show this comparison in the case of the spin $S=1$ for the nematic ($h=2.6$)  and supernematic ($h=4.29$) phases (Fig.~\ref{fig:energy_comparison}(a-b)) for which the unit cell is made of only three different sites. The environment was calculated using the MPS-based approach, with the couple $\{D,D_c\}$, where $D_c$ is the environment bond dimension, $\{3,18\}$, $\{4,23\}$ and $\{5,30\}$ for the nematic phase and $\{3,29\}$, $\{4,29\}$, $\{5,31\}$ and $\{6,30\}$ for the supernematic phase.

\section{Reduced magnetization}
\setcounter{figure}{0}   

\begin{figure*}
\includegraphics[trim=0 0 0 0,clip ,width=1\textwidth]{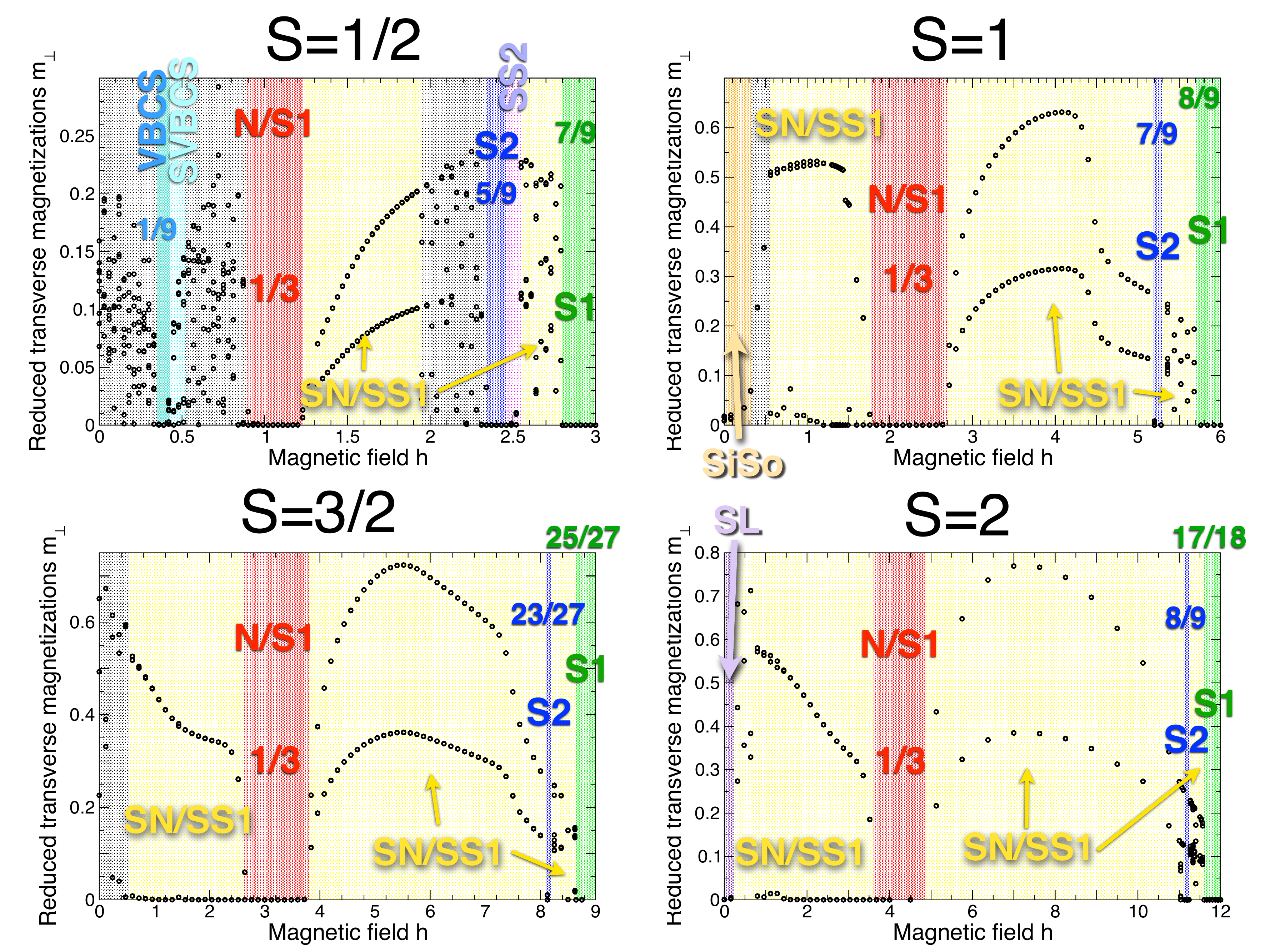}
\caption{
[Color online] Reduced transverse magnetizations (normalized w.r.t. its value at saturation) versus the external magnetic field for (a) $S=\frac{1}{2}$, (b) $S=1$, (c) $S=\frac{3}{2}$ and (d) $S=2$, with bond dimension $D=10$. 
The plateau phases (in red, light blue, dark blue and green) and the gaped phases for spin $S=1$ and $S=2$ are characterized by a zero local transverse magnetization for every site.
Yellow regions are compressible supernematic or supersolid phases. Although the local magnetizations are non-zero, the total transverse magnetization is zero $\vec{m}_{\perp}^{\text{tot}}=\sum_{\mu}\vec{m}_{\perp}^{\mu}=\vec{0}$. Grey regions could not be fully characterized within our approach as show the transverse magnetization behavior. The phases are labelled according to the notations in Table I and the colors correspond to the Fig.2.}
\label{fig:m_xy}
\end{figure*}

As we apply an external magnetic field, the systems go through several phases. In order to identify these different phases, we calculcate local expectation values using the procedure describes in the \textit{Supplemental Material}~\cite{spin1pd}. First, we compute the reduced longitudinal magnetization, as presented before, on each site $m_z^{\mu}=\langle S_z^{\mu}\rangle/S$, where the label $\mu\in [1,9]$ runs over the nine sites. The average of the magnetization along the magnetic field $m_z=\sum_{\mu}m_z^{\mu}/9$ is especially useful for identifying the several magnetization plateaus. The local reduced magnetizations along the magnetic field together with the local reduced magnetizations transverse to the magnetic field $m_{\perp}^{\mu}=\sqrt{\langle S_x^{\mu}\rangle^2+\langle S_y^{\mu}\rangle^2}/S$ give information about the symmetries of the phases. The transverse magnetization is shown in Fig.~\ref{fig:m_xy}. The $SU(2)$-symmetry breaking is characterized by a non-zero local magnetization in the transverse plane. Since the plateau phases preserve this spin symmetry, all the local transverse magnetization are zero, as shown in Fig.~\ref{fig:m_xy}, for every spins. Moreover, in every phases, but the gray regions, the global $SU(2)$-symmetry is preserved, which is characterized by $\vec{m}_{\perp}^{\text{tot}}=\sum_{\mu}\vec{m}_{\perp}^{\mu}=\vec{0}$

In addition, we calculate the eigthteen local energies $e_{i,j}=\langle\mathbf{S_i}\cdot\mathbf{S_j}\rangle$ for $i$ and $j$ nearest neighbours. This density is pictured in the Fig. 1(c-e) of the main paper and is necessary to characterize the Valence Bond Crystal phase.

